# The superconducting and magnetic states in $RuSr_2GdCu_2O_8$, based on the magnetic, transport and magneto-caloric characteristics

Piotr W. Klamut *, Tomasz Plackowski

*Institute of Low Temperature and Structure Research, Polish Academy of Sciences, P Nr 1410, 50-950 Wrocław 2, Poland*

**Abstract**

The article discusses selected properties of the non- and superconducting polycrystalline samples of $RuSr_2GdCu_2O_8$ and comments the consequences of introducing insignificant sub-stoichiometry of Ru into the nominal formula. The magneto-resistive and the magnetic characteristics are interpreted in favour of the formation of the intrinsically inhomogeneous superconducting phase, which seems to be stabilized along with the structural modifications likely enhanced with the modification of starting stoichiometry. The specific heat data reveals the shift of temperature of the magnetic ordering $T_m$, suggesting the dilution in magnetic sublattice of the Ru moments. The measurements of the magnetic field dependences of the isothermal magnetocaloric coefficient $M_T$ show that there is no gain in magnetic entropy in a broad range of the accessed fields and temperatures. Whereas the multi-component character of the probed magnetic system precludes from concluding on the ground state for the Ru ordering, the maximum in $M_T(H)$ which occurs at weak magnetic fields for temperature vicinity of $T_m$ may reflect dominance of the ferromagnetic type interactions with a constrained correlation range. The literature explored models for the Ru magnetic ordering and possible phase separation in $RuSr_2GdCu_2O_8$ are brought into the discussion.

Ruthenate-cuprates form a separate class in the family of complex copper oxides, which is distinguished by the magnetically active sublattice of Ru built of the Ru-O structural layers. There have been few types of the ruthenate-cuprate crystal structures identified, which differ by length of their layered unit cell, with structural blocks characteristic for the HTSC cuprates. Manifold properties of these compounds were covered in numerous reports and also in several review texts complementing each other for emphasising different aspects of the field [1-6 and references therein]. The 1212-type structure of the $RuSr_2GdCu_2O_8$ resembles that of the $REBa_2Cu_3O_7$ superconductor, in which the so called chain positions of Cu ions are replaced with Ru in then full octahedral coordination with oxygens. The Sr alkaline earth metal assumes analogous structural positions to Ba in the $REBa_2Cu_3O_7$. The 1212-type M-cuprates were reported with several transition metal ions M=Ru, Cr, Ir, and several rare earth (RE) elements [7], for which so far only the ruthenate-cuprates were found superconducting. The magnetic ordering in the Ru sublattice in $RuSr_2GdCu_2O_8$ extends up to $T_m \approx 135$ K whereas the paramagnetic sublattice of $Gd^{3+}$ ions orders anti-ferromagnetically below 2.5 K [8]. Since $RuSr_2GdCu_2O_8$ was reported as superconducting with $T_c$ as high as approximately 50 K, it became very interesting to investigate simultaneous accommodation and mutual influence of the superconducting order parameter and the Ru magnetism. In the temperature range below $T_m$, the field dependencies of the *dc* magnetisation become hysteretic in resembling behaviour of the magnetisation of a weak ferromagnet. The temperature dependence of the internal field associated with the Ru ordered state below $T_m$ was established in zero field µSR (muon spin rotation) measurements, with no change in field values detected upon entering the temperature range of the superconducting phase [9].

The magnetic ground state of the Ru sublattice in $RuSr_2GdCu_2O_8$ was approached with several different models, which were based in several different experiments. Due to the lack of suitable single crystals, most were so far based in investigation of the sintered polycrystalline samples, which for some experiments contributes complexity to interpretation. Early neutron powder diffraction (NPD) experiments [8,10] concluded the G-type antiferromagnetic order of Ru with easy axis pointing along the *c* direction and associated the ferromagnetic diffractive maxima to the effect of field dependent canting of the $Ru^{5+}$ magnetic moments. The field dependent magnetic diffraction analyzed in the following $Gd^{160}$ NPD experiment for assuming the above, pointed to the necessity of rather fast increase of the canting angle, with more than 75˚ at magnetic field values above 1.5 T [11]. Measurements of the zero-field and field dependent Ru-NMR spectra also provided arguments for the AFM ground state in the Ru sublattice. Observed field dependences of the resonance frequencies led to propose type-I AFM structure built of the ferromagnetic *a-b* planes, and spin flop transition to ferromagnetism involving field induced rotation of the in-planes magnetisation vectors, which seems to be allowed for expectance of substantial magnetic anisotropy of the Ru magnetic lattice [12]. Several experiments also evidenced two valence states of the Ru ions in the $RuSr_2GdCu_2O_8$, $Ru^{5+}$ and $Ru^{4+}$ [12-15]. Most recently communicated investigation of the magnetic state by the resonant X-ray diffraction with data collected for the single crystal of superconducting $RuSr_2GdCu_2O_8$, reconciles some of the previous hypotheses with an observation of the G-type structure with the Ru magnetic moments pointing along the low symmetry direction, then with direction of the ferromagnetic in-plane component alternating in stack along the *c*-axis [16]. Last but not least, we note the analyses of the temperature and the field dependencies of macroscopic characteristics for the superconducting and magnetic phases in ceramic samples of $RuSr_2RECu_2O_8$ (RE=Gd, Eu) led to postulate of the material inherent spatial phase separation to an ensemble of antiferromagnetic/superconducting and ferromagnetic/non-superconducting domains within the single crystallites [1].

The superconducting samples of $RuSr_2GdCu_2O_8$ were reported with wide spread of $T_c$ and with maximum value of the transition onset temperature 52-55 K [18, 19], latter also for the first reported single crystals. The nominally same stoichiometric $RuSr_2GdCu_2O_8$ was, however, also found non-superconducting [19, 20, 21]. Figure 1, after [21], shows the temperature dependences of the *ac* susceptibility collected for three ceramic samples of the $RuSr_2GdCu_2O_8$, two of which (B and C) were obtained through additional annealing (1060 ˚C in oxygen) of initially non-superconducting compound (A). The difference between samples B and C is in slow *vs.* fast cooling, i.e. the annealing converted the material superconducting (sample B) and its diamagnetic shielding was secured only for the slow cooled (sample C). The non-superconducting sample A was synthesised in the solid state reaction performed at 935 ˚C in flow of 1%$O_2$/Ar (Ar balance gas), i.e. at substantially lower temperature than 1050-1060 ˚C which is required for synthesis of this phase in flowing oxygen. Modified conditions were based in finding of the approximately 20 K wide plateau in the mass-temperature TGA data collected at 1%$O_2$/Ar, indicating the range of thermo-dynamical stability for the 1212-type phase. Neither the phase purity of these samples (single phase powder XRD patterns) nor their overall oxygen content, were found distinguishable. The temperature dependencies of the crystal lattice parameters for the non-superconducting sample A and the superconducting sample B sample were accessed in the synchrotron X-ray diffraction experiment [11], which revealed insignificant decrease of the Cu-O2 distance and larger increase of the Cu-O3 distance for the superconducting material, with latter difference becoming more evident below temperature of magnetic ordering of the Ru sublattice ($T_m \approx$ 133-136 K). The O3 refers to the crystallographic position of the oxygen atom in between the Ru and Cu (i.e. the apical oxygen in the $RuO_6$ octahedron) and O2 to the oxygens in the $CuO_2$ plane (the Cu-O2-Cu angle was established in that experiment at approximately 168.6˚-169.1˚ for both the NSC and SC samples). The differences translated to a slightly larger *c/a* ratio for the superconducting sample (3.0145 *vs.* 3.0163 at 10 K) which was found to increase at $T_m$ for both samples. The Williamson-Hall plots constructed for the (hk0) and (00l) synchrotron XRD reflections showed a difference in the slope, indicating a higher degree of the lattice strain in the non-superconducting compound [11].

In Ref. 22 we comment more extensively on the synthesis of two samples with slightly altered nominal stoichiometry, $RuSr_2GdCu_2O_8$ and $Ru_{0.98}Sr_2GdCu_2O_8$, whose selected properties will also be discussed here. The approach was to modify the nominal formula for one of to be compared samples introducing slight deficiency of Ru, and for keeping the same reaction parameters during simultaneous synthesis, to end up with the materials whose properties may be compared in investigation of the possible modification of the phase. This acquired additional context in recently reported studies of the local crystal structure of superconducting $RuSr_2GdCu_2O_8$ by means of the high resolution electron microscopy [23]. It was suggested that superconducting phase in this material may originate within the nanometre range domains distinguished by doubled *c* axis parameter for replacing every second Ru-O layer with so called Cu-O chain layer, in resemblance to the related structure of YBCO superconductor.

The solid state synthesis of two considered samples was done in flowing oxygen at 1060 ˚C, which corresponds to the conditions often reported for synthesis of $RuSr_2GdCu_2O_8$, albeit which are close to the melting point for the phase. Even the final reaction was performed out of mixture of the preformed precursors (in intermediate step the previously calcined substrates were reacted to the mixture of the $Sr_2RuGdO_6$ and rest oxides, predominantly $Cu_2O$), one should take into account a rather high sublimation rate for the Ru rich oxides at the high temperature oxidative conditions, which appear to be required for the final synthesis of the XRD single-phase and superconducting material. The samples then may be set vulnerable to local modifications of their stoichiometry with the candidate Ru/Cu sites in the 1212-type phase. The inhomogeneous structural modifications may occur preferentially in the micro-regions with slightly altered stoichiometry also for nominally stoichiometric material, even associating with trace melting for larger local off-stoichiometry [20,22]. Both samples considered here were simultaneously cooled at the slow rate 1°C/min, which would compare a differently prepared sample B in figure 1. The influence of rate of cooling, which determines the diamagnetic shielding is still difficult to interpret. It would be intuitive to associate it with the uptake of oxygen along the superconducting paths in the material, there is however a possibility of structural ordering processes, which at present could not be verified experimentally. We note that recent investigation of the synthesis associated sublimation processes provided for even macroscopic inhomogeneity of the Ru content in the bulk vs. surface of the ceramic sample of $RuSr_2GdCu_2O_8$ [24] (extensive commentary on the synthesis issues for ruthenocuprates was presented by this group in [2, 25]). In this comparative study, by including the sample with subtle modified nominal stoichiometry, we set an alternative approach, which aims at induced differences in the material properties for promoting a different extent of the processes associated with the Ru/Cu off-stoichiometry.

Figure 2 presents the magnetic field dependencies of the temperatures, which were chosen to be characteristic for the superconducting transitions in the $RuSr_2GdCu_2O_8$ and $Ru_{0.98}Sr_2GdCu_2O_8$ samples, and compares them to analogous defined temperatures read of the resistivity-temperature dependence of the $Ru_{0.5}Sr_2GdCu_{2.5}O_{8-d}$ . The latter, structurally related superconductor has higher $T_c$ and shows no signatures of the Ru magnetism in its then paramagnetic normal state [26,21]. These characteristics were more extensively discussed in [27] together with observed anomalous increase of conductivity in the magnetic field ahead of superconducting transition, which we linked to complex response of the constrained dimensionality superconducting phase in presence of the compound's magnetism (see also ref. 28). Figure 3 presents the corresponding temperature dependencies of the *ac* susceptibility measured at two small amplitudes of the *ac* field - 1 Oe and 0.1 Oe. Increase of the $T_c$ in the $Ru_{0.98}Sr_2GdCu_2O_8$ compared to the $RuSr_2GdCu_2O_8$ (figure 2 and 3) might suggest only a modification of the effective charge doping in the superconducting phase, common for shifting $T_c$ in the superconducting cuprates. It is, however, of the other differences, as is the different slope of the magnetic field dependence of $T_c^{on}$ (figure 2) and shift of the magnetically measured onset of the transition in response to different amplitudes of the *ac* field for its limiting small values (figure 3), that the superconducting phase in the $RuSr_2GdCu_2O_8$ appears rather spatially confined with characteristic length scale of the order of magnetic penetration depth $\lambda$. It would then seem natural that the phase likely develops along with these structural modifications, which may associate or reflect from nominally decreased Ru/Cu ratio. Such scenario would also accommodate differently prepared $RuSr_2GdCu_2O_8$ which remains not superconducting, and may link superconductivity with substantially higher critical temperatures found for related $Ru_{1-x}Sr_2GdCu_{2+x}O_{8-d}$, ($T_{c,max}$=72 K for the x=0.5 and oxygen annealed phase) which for larger x was stabilised in course of the high pressure annealing [26]. Slight increase of the superconducting $T_c$ was also observed in nominal $Ru_{0.9}Sr_2GdCu_{2.1}O_8$ *vs.* nominally 1212 stoichiometric sample, reported in [29] together with the effect of prolonged time of oxygen annealing leading to an increase of $T_c$.

Beyond the ceramic samples of $RuSr_2GdCu_2O_8$, the issue of intrinsic spatial homogeneity of its superconducting phase was accessed in one of the first experiments performed on single crystals [30], which evidenced the Josephson type coupling along the *c* axis, so and the effectively layered structure built of interlaced *SC-I-SC* slabs. No peculiar features in the I-V characteristics were found in that experiment to be associated with the Ru magnetism. Note, that qualitatively similar Josephson characteristics have been observed for several other underdoped HTSC cuprates with sufficiently large structural anisotropy (Bi-2212, Tl-2223, [31] and therein) so that the feature remains nonspecific to the ruthenocuprate in the HTSC family [30]. For the single crystals of $RuSr_2GdCu_2O_8$, similar to those communicated in [30] and [32], recently reported resonant X-ray diffraction yet concludes [16] the presence of ferromagnetic component along the $RuO_2$ layers with its alternating direction along the *c* axis, essentially in agreement with model proposed for early NMR results [12]. What may turn relevant from perspective of eventual local alternations in the 1212-type structure of the crystals, is that all of the so far investigated crystals show comparatively high values of the superconducting $T_c \approx$ 45-60 K and substantially lowered temperature of the Ru magnetic transition ($T_m$ =102 K was inferred from the temperature dependence of the intensities of two resonant RXD magnetic reflections as well as from the *dc* magnetisation data [16], in ref. [16] the upturn of magnetic susceptibility of the crystal in wide temperature range of approximately 40 K above that $T_m$ was proposed to originate from non-compensated ferromagnetic moments to appear at local structural modifications in consequence of much weaker magnetic interactions across the $RuO_2$ layers in the Ru sublattice). Possible modified distribution of the Cu, Ru or O ions in the crystal in respect to the polycrystalline materials was brought in [16] for a likely reason of observed differences. It may be commented after [2] that for such crystals grown at conditions of equilibrium of the liquid and solid phases the homogeneity range may become restricted and in result of the temperature dependent process the crystal may turn be not homogeneous, perhaps for that even leaving superiority to polycrystalline material.

Figure 4 presents the temperature dependences of the specific heat for the considered samples $Ru_{0.98}Sr_2GdCu_2O_8$ and $RuSr_2GdCu_2O_8$, after [6]. Notice the small shift of $T_m$ toward lower temperatures for the $Ru_{0.98}Sr_2GdCu_2O_8$ . The same is seen in the *ac* susceptibility data (figure 1), the specific heat, however, proves bulk character of the feature. Figure 5 presents the magnetic field dependence of $T_m$ established from the specific heat-temperature dependences collected for the $RuSr_2GdCu_2O_8$ , compared in that figure with the characteristic onset temperatures for magnetically ordered state inferred from the magneto-resistivity data measured for both the $Ru_{0.98}Sr_2GdCu_2O_8$ and $RuSr_2GdCu_2O_8$ samples. The chosen characteristic temperatures correspond to the initial increase of conductivity, which signals the Ru magnetic order in the ruthenate-cuprate, the effect interesting by itself for associating the ordering with a conduction channel in the Ru sublattice. Despite rather large temperature errors (dotted lines), which come with determination of the $T_m$ from transport, we note that sufficient agreement holds between such chosen temperatures and the $C_p(T)$ determined characteristic temperatures, which may be taken for proving the bulk character of the accessed features.

It may be noted that since the effect of magnetic field is to increase $T_m$, it would be unusual for assuming the antiferromagnetic type of interactions as leading to spatial long range order at $T_m$. Note that a considerable shift of $T_m$ is observed for comparatively weak fields – in consideration of the role of magnetic anisotropy, for already proposed models involving the field induced rotation of planar easy magnetisation axes [12] or progressing the spatial ordering with long range but weak planar dipolar interactions (formulated in [33] for the related structure of 1222-type ruthenate-cuprate), remains intuitive for such data. Then, primary role of the magnetic field in temperature vicinity of the transition would be in locking in the 3D long range spatial order. Regarding the $T_m(H)$ dependences presented in figure 5, note that the different slope for the $Ru_{0.98}Sr_2GdCu_2O_8$ and $RuSr_2GdCu_2O_8$ may result from nonlinear correspondence between the extent of introduced dilution in magnetic sublattice of Ru and its effect on the $T_m(H)$ (should not assume weakly superconducting $RuSr_2GdCu_2O_8$ sample for referencing the response of ideal magnetic lattice of the Ru moments in the 1212-type structure). Last but not least, we note that for the $RuSr_2GdCu_2O_8$ one deals with partly itinerant 4d electrons on the Ru lattice so there seems no *a priori* reason for excluding band polarisation effects for its magnetism (note a similarity in conductivity response at the ordering temperatures for the $RuSr_2GdCu_2O_8$ and the $SrRuO_3$ itinerant ferromagnet).

In the following part we comment on approaching the magnetic state in the $RuSr_2GdCu_2O_8$ in the measurements of the isothermal magnetocaloric coefficient $M_T$, which has an advantage of directly probing the magnetic entropy changes and as a thermodynamic measure constrains to bulk features of the samples. Measurements of the $M_T$ were performed in a heat flow calorimetric setup described in [34] at the conditions of constant temperature and at constant rate of sweeping magnetic field. Then,

$$M_T \equiv \frac{dq}{dB} = \frac{-\dot{q}}{\dot{B}} = \frac{-U}{A\dot{B}}$$

where *q*-heat flux, B-magnetic induction, U- voltage generated by the heat flow meter, A-sensitivity parameter.

The measurement accesses changes of magnetic entropy of the system according to relation:

$$\Delta S_{mag}(T,B) = -\int_0^B \left(\frac{M_T}{T}\right) dB,$$

and the $M_T$ relates to magnetization with the formula:

$$M_T = -T \left(\frac{\partial M}{\partial T}\right)_B.$$

Figure 6 presents the set of the magnetic field dependencies of the $M_T$ measured for the $RuSr_2GdCu_2O_8$ sample, which properties were commented for measurements of the *ac* susceptibility and resistivity. Note that for a simple antiferromagnet, one would expect the negative values of $M_T$ for field driven increase of its magnetic entropy. For the $RuSr_2GdCu_2O_8$ data it should be noted that the compound contains few magnetic sublattices, for which the sublattice of comparably large $Gd^{3+}$ magnetic moments ($\mu_{eff}(Gd^{3+})$=7 $\mu_B$ [8]) remains paramagnetic in a whole range of the accessed temperatures. Predominant contribution of the $Gd^{3+}$ paramagnetism to $M_T$ is seen in figure 6 for temperatures sufficiently above $T_m$. The extra increase of the $M_T(H)$ with maximum at finite magnetic fields observed for temperatures in a vicinity of $T_m$ suggest presence of the ferromagnetic interactions associated with the ordering at $T_m$. For the magnetic system the field dependence of the magnetocaloric coefficient at the transition temperature should strongly diverge at zero field [35], so and the maximum observed for finite field values may suggest no spontaneous long range order in favor of the field induced ferromagnetism. Then, only finite increase of the $M_T$ may signal constrained correlation length for responsible magnetic interactions. These conclusions would be justified for the literature proposed inherent inhomogeneity of the Ru magnetic system [1] as well as for above discussed role of the magnetic field in inducing long range ferromagnetic order in the Ru sublattice. For further interpretation of the $M_T(H)$ data one should, however, take into account the possible polarization effects between the compound's separate magnetic sublattices, which for the $RuSr_2GdCu_2O_8$ involve large magnetic moments of the $Gd^{3+}$ions [36]. It would then seem desirable to extend such measurements for other rare earth based ruthenate-cuprates, including the high pressure synthesized $RuSr_2YCu_2O_8$.

In summary, the text comments on several characteristics of the superconducting and magnetic transitions in the samples of $RuSr_2GdCu_2O_8$, most possibly coupled with the postulated local scale peculiarities in the compound's structure. It would be tempting to elucidate if the constrained dimensionality of the superconducting phase in the $RuSr_2GdCu_2O_8$ may meet with the sufficient material perfecy to become relevant in investigation of the quasi 2D regime of superconductivity in cuprates.

**Acknowledgment**

This research was partly financed by the Polish Ministry of Science and Higher Education research project funding for the years 2007–2010. Most of the presented results conforms the author's presentations given at the XIV National School for Superconductivity in Ostrów Wielkopolski, in October 2009, and at the 9$^{th}$ International Conference on Materials and Mechanisms of Superconductivity in Tokyo, in September 2009.

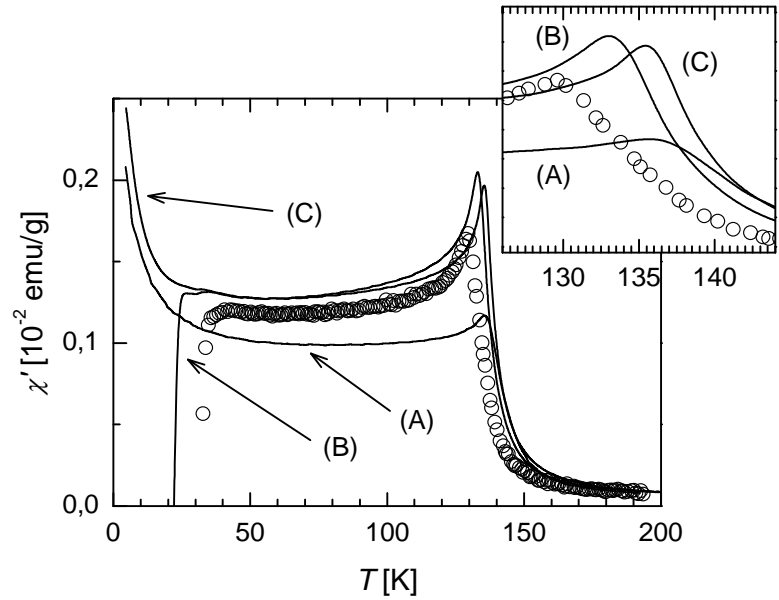

Figure 1. Temperature dependencies of the real component of the *ac* susceptibility ($H_{ac}$=1 Oe, $f$=200 Hz) for three samples obtained from the initially non-superconducting (NSC) $RuSr_2GdCu_2O_8$: *(A)→ (B) → (C)* represents the sequence of annealing: (A) as grown in 1%$O_2$/Ar at 935˚C, (B) after subsequent annealing at 1060 ˚C in $O_2$ for 140 hours with slow cooling, (C) same to B but quenched from $O_2$/1060˚C. Open circles: sample obtained afer annealing of the NSC $RuSr_2GdCu_2O_8$ in 600 bar of oxygen at 1100 ˚C, slow cooled.

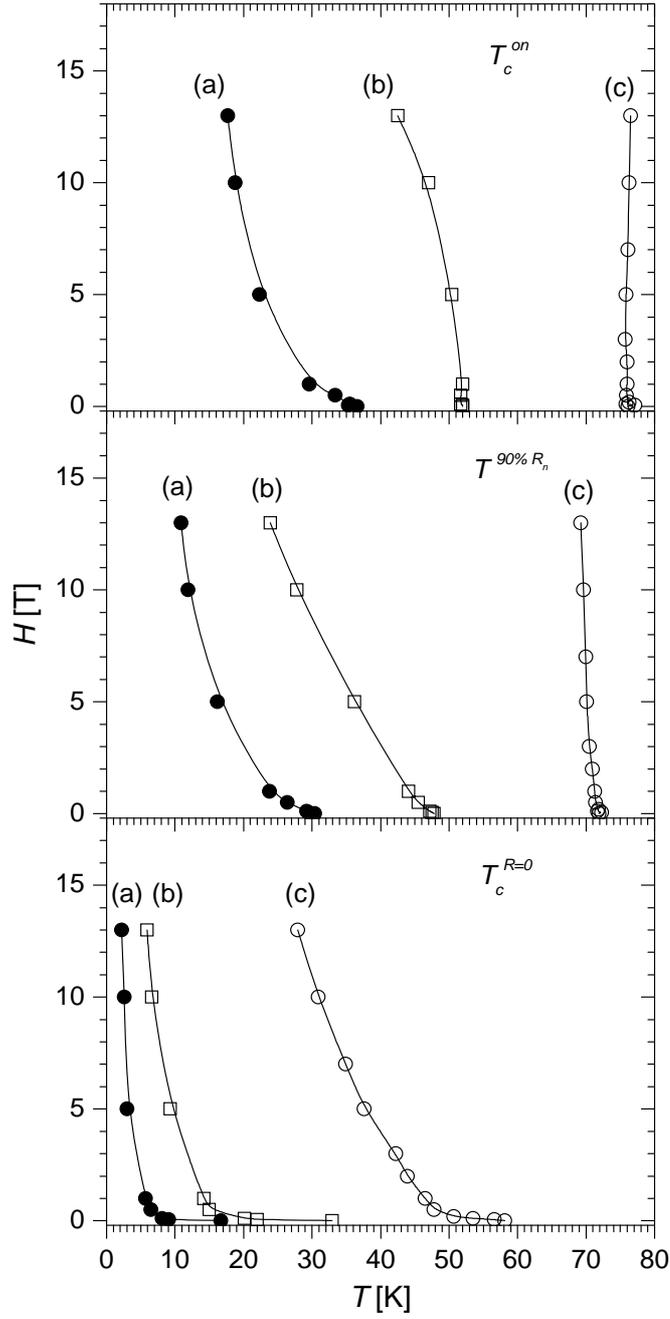

Figure 2. The magnetic field dependencies of the temperatures characteristic for the resistive superconducting transitions in the samples: (a) $RuSr_2GdCu_2O_8$, (b) $Ru_{0.98}Sr_2GdCu_2O_8$ and (c) $Ru_{0.5}Sr_2GdCu_{2.5}O_{8-d}$ (oxygen annealed for small $d$). Upper panel: $T_c^{on}$ – temperature of the onset of transition, middle panel: $T_c^{R=0}$ – uppermost temperature of the zero resistivity state, lower panel: $T^{90\% R_n}$ – temperature corresponding to the 10% decrease of resistivity relative to its value at $T_c^{on}$.

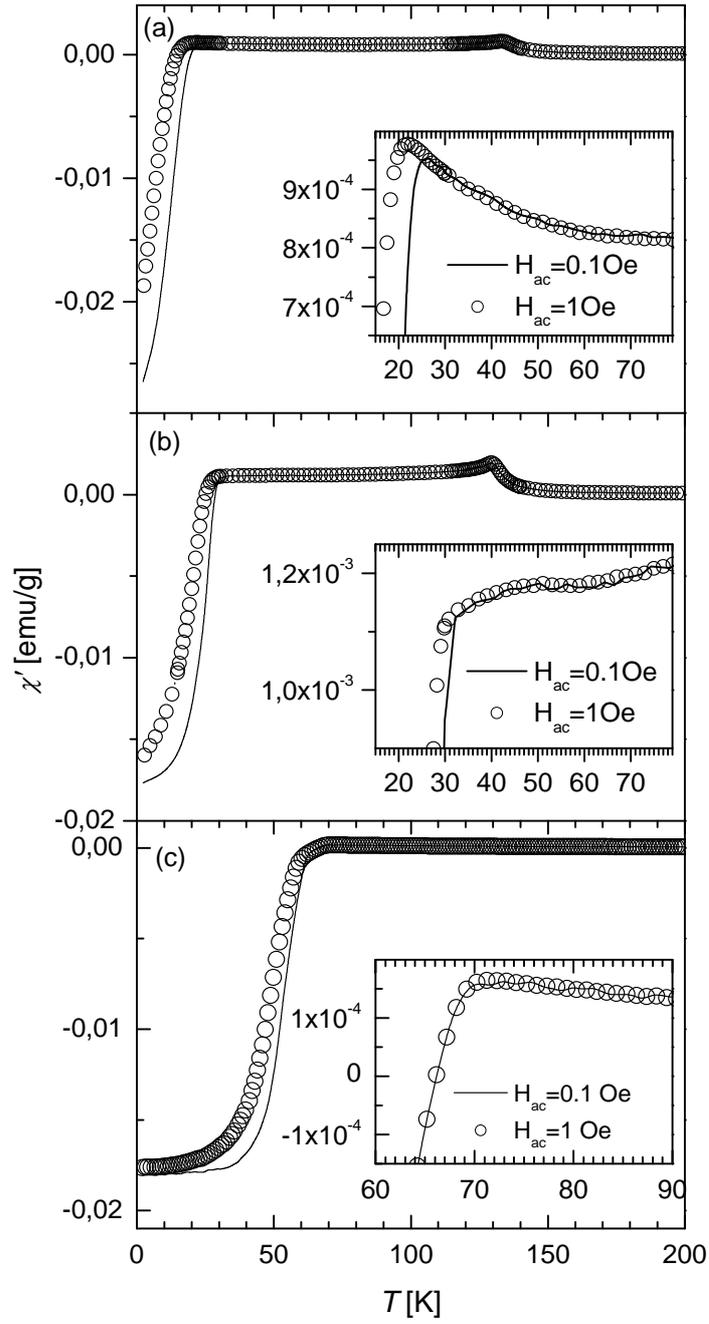

Figure 3. Temperature dependences of the real component of the *ac* susceptibility for (a) $RuSr_2GdCu_2O_8$, (b) $Ru_{0.98}Sr_2GdCu_2O_8$, (c) $Ru_{0.5}Sr_2GdCu_{2.5}O_{8-d}$ samples. $f = 1$ kHz.

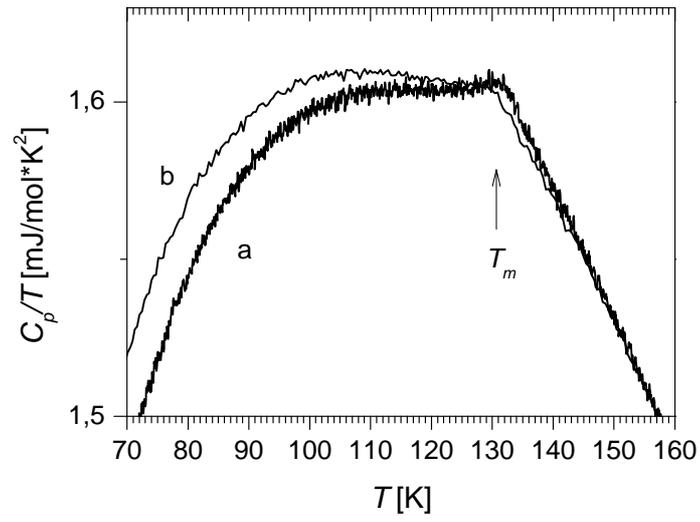

Figure 4. Plot of $C_p/T$ vs. $T$ for (a) the RuSr$_2$GdCu$_2$O$_8$ and (b) Ru$_{0.98}$Sr$_2$GdCu$_2$O$_8$ sample. Note slight shift of the characteristic temperature associated with $T_m$.

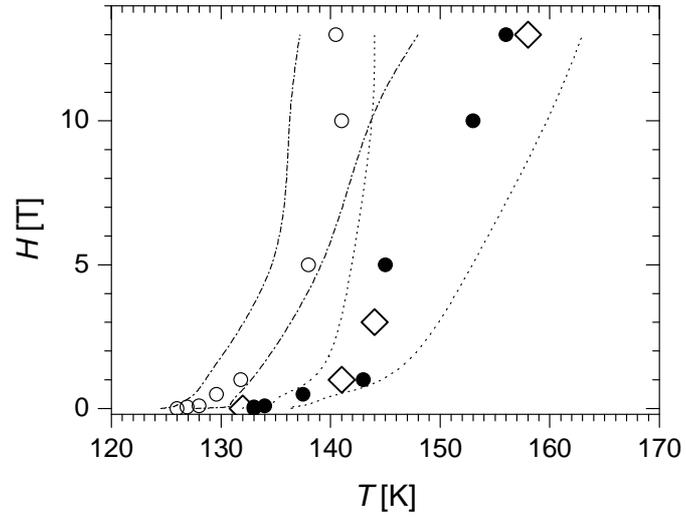

Figure 5. The magnetic field dependencies of the temperatures characteristic for increase of conductivity in a vicinity of $T_m$ measured for simultaneously prepared $RuSr_2GdCu_2O_8$ (closed circles) and $Ru_{0.98}Sr_2GdCu_2O_8$ (open circles). Dotted and dash-dotted lines show estimated uncertainty of determination of these temperatures from the $\rho(T)$ data. Open diamonds show the $T_m$ associated temperatures in $RuSr_2GdCu_2O_8$ as they were determined from the temperature dependencies of the specific heat.

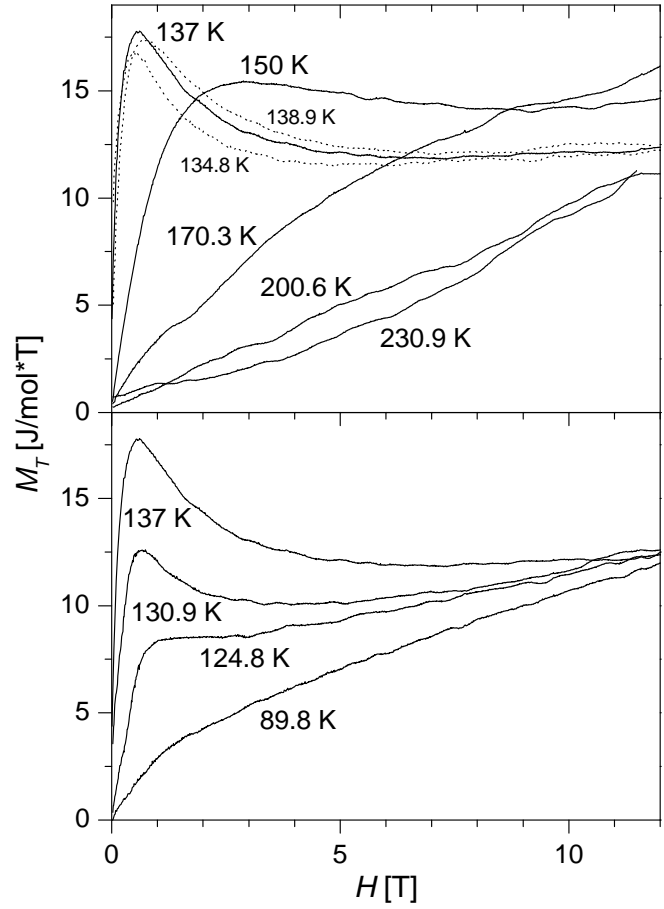

Figure 6. Magnetic field dependencies of the isothermal magneto-caloric coefficient measured for the $RuSr_2GdCu_2O_8$ sample at several temperatures above (upper panel) and below (lower panel) $T_m$. The maximal height of the maximum at the field of approx. 0.5 T (note the surrounding dependences collected at 138.9 K and 134.8 K) was found for 137 K, which temperature matches the $T_m$(H=0.5 T) determined from the specific heat data for this sample (see figure 5).[28]